\documentclass[arguments]{aastex631}

\usepackage{amsmath}
\usepackage{gensymb}
\usepackage{textcomp}
\makeatletter
\newcommand{\mathleft}{\@fleqntrue\@mathmargin0pt}
\newcommand{\mathcenter}{\@fleqnfalse}
\makeatother

\usepackage{CJK}

\begin{document}

\begin{CJK*}{UTF8}{gbsn}
\title{Langmuir Wave Excitation in Solar-wind Magnetic Holes}

\author[0009-0003-9856-5949]{Jingting Liu (刘 婧婷)}
\affiliation{Mullard Space Science Laboratory, University College London, Dorking RH5 6NT, UK}

\author[0000-0002-0497-1096]{Daniel Verscharen}
\affiliation{Mullard Space Science Laboratory, University College London, Dorking RH5 6NT, UK}

\author[0000-0002-2576-0992]{Jesse Coburn}
\affiliation{Mullard Space Science Laboratory, University College London, Dorking RH5 6NT, UK}
\affiliation{The Blackett Laboratory, Imperial College London, London, SW7 2AZ, UK}

\author[0000-0003-3623-4928]{Georgios Nicolaou}
\affiliation{Mullard Space Science Laboratory, University College London, Dorking RH5 6NT, UK}

\author[0000-0001-7019-5905]{Xiangyu Wu (吴 翔宇)}
\affiliation{Mullard Space Science Laboratory, University College London, Dorking RH5 6NT, UK}

\author[0000-0001-7431-5759]{Wence Jiang (蒋 文策)}
\affiliation{State Key Laboratory of Solar Activity and Space Weather, National Space Science Center, CAS, 100190, Beijing, China}

\author[0000-0002-7638-1706]{Oreste Pezzi}
\affiliation{Istituto per la Scienza e Tecnologia dei Plasmi, Consiglio Nazionale delle Ricerche (ISTP-CNR), 70126 Bari, Italy}

\author[0000-0002-5272-5404]{Francesco Pucci}
\affiliation{Istituto per la Scienza e Tecnologia dei Plasmi, Consiglio Nazionale delle Ricerche (ISTP-CNR), 70126 Bari, Italy}

\author[0000-0002-0282-2978]{Matteo Zuin}
\affiliation{Consorzio RFX (CNR, ENEA, INFN, Universit\`a di Padova, Acciaierie Venete SpA), 35127 Padova, Italy}
\affiliation{Istituto per la Scienza e la Tecnologia dei Plasmi del CNR,  35127 Padova, Italy}

\author[0000-0002-5982-4667]{Christopher J. Owen}
\affiliation{Mullard Space Science Laboratory, University College London, Dorking RH5 6NT, UK}

\author[0000-0002-6287-3494]{Hamish Reid}
\affiliation{Mullard Space Science Laboratory, University College London, Dorking RH5 6NT, UK}




\begin{abstract}
Magnetic holes are structures commonly observed in various space plasma environments throughout the solar system, including the solar wind. These structures are characterized by a localized decrease in magnetic field strength, coincident with an increase in plasma density.
Previous observational studies in the solar wind link the presence of Langmuir waves to magnetic holes, suggesting a strong correlation between these phenomena.
We develop a model based on magnetic-moment conservation and its violation to explain the excitation of Langmuir waves in magnetic holes. Our model illustrates that magnetic holes induce changes in the electron velocity distribution function that emit electrostatic Langmuir waves due to the bump-on-tail instability. Using data from the Solar Orbiter spacecraft, we provide a comprehensive analysis of this process and test our predictions with observations. The consistency between the model and observations indicates that our proposed process is a viable mechanism for producing Langmuir waves in magnetic holes in the solar wind.
\end{abstract}



\section{Introduction} \label{sec:intro}

Magnetic holes (MHs) are localized regions in space plasmas in which the magnetic field strength decreases significantly, often by more than 50\%. These structures were first identified in the solar wind \citep{turner1977magnetic}. Since then, they have been observed across a variety of space environments, including planetary magnetosheaths \citep{tsurutani1982lion, lucek1999mirror}, the Earth's cusp region \citep{shi2009spatial}, planetary magnetotails \citep{ge2011case, sun2012cluster}, and cometary environments \citep{russell1987mirror}. MHs vary in spatial scale, ranging from a few electron gyro-radii $\rho_\mathrm{e}$ to thousands of proton gyro-radii $\rho_i$ \citep{winterhalter1994ulysses, sperveslage2000magnetic}. Several theories exist to explain the formation of MHs, including mirror-mode instabilities \citep{tsurutani1982lion,zhang2008characteristic,xiao2014plasma}, solitary waves \citep{burlaga1978interplanetary}, interchange instabilities \citep{lapenta2011self}, Kelvin-Helmholtz instability \citep{arro2023generation}, and turbulence \citep{arro2024largescale}. MHs exhibit a higher density than the surrounding plasma in order to maintain pressure balance.

One of the ways in which MHs influence the plasma is through their interaction with waves: various types of waves are observed within MHs, including whistler waves \citep{huang2020excitation, yao2019waves,jiang2022whistler}, electrostatic solitary waves \citep{yao2019waves}, and electron cyclotron harmonic waves \citep{zhang2017kinetics}.  These waves are absorbed or emitted through resonant interactions with different instability mechanisms linked to the trapped electron and ion populations inside the MHs \citep{jiang2024}.

Langmuir waves (LWs) are electrostatic plasma waves with wave frequencies near and above the electron plasma frequency $\omega_{\mathrm {pe}}$. They are of particular interest for their fundamental role in converting energy through wave--particle interactions with electrons in weakly collisional plasmas such as the solar wind \citep{herr2016introduction}. While interplanetary LWs are often associated with type II and type III radio bursts \citep{cairns1986source, robinson2000bidirectional, pulupa2020statistics}, a significant portion of LWs in the solar wind originate from MHs rather than radio bursts \citep{lin1996langmuir, boldu2023langmuir}. LWs occur more frequently in association with magnetic than in the surrounding solar wind \citep{boldu2023langmuir}.

The solar wind electron velocity distribution function (VDF) typically consists of a Maxwellian core, an isotropic suprathermal halo, and a field-aligned suprathermal strahl \citep{feldman1975solar, verscharen2019multi}. The electron strahl is a skewed extension of the core population in the direction parallel or anti-parallel to the magnetic field. It typically does not form a bump-on-tail configuration with a region of positive velocity gradient of the electron distribution. Hence, the strahl population is unlikely to generate electrostatic bump-on-tail instabilities in the ambient solar wind \citep{verscharen2019self,horaites2018stability,schroeder2021stability}. 
 
We propose that MHs can modify the strahl electron configuration in such a way that a local bump-on-tail distribution develops within the MH. This bump-on-tail configuration then drives LWs through resonant wave--particle interactions.
Our model is based on the conservation of adiabatic invariants. 

\section{A Model for Langmuir wave generation in MHs} 

We first briefly introduce the bump-on-tail instability of LWs and the conservation of the magnetic moment $\mu$ conceptually. We then discuss the breaking of $\mu$-conservation and present our model for the modification of the electron VDF due to the breaking of $\mu$-conservation in MHs.

\subsection{Bump-on-tail instability}

The bump-on-tail instability arises when the local gyrotropic particle velocity distribution function $f(v_{\perp},v_{\parallel})$ develops a beam signature in such a way that $ \partial f /\partial v_{\parallel} > 0$ at $v_{\parallel}>0$ or that $ \partial f /\partial v_{\parallel} < 0$ at $v_{\parallel}<0$, where $v_{\perp}$ and $v_{\parallel}$ denote the perpendicular and parallel velocity components with respect to the local magnetic field direction. For the sake of simplicity but without loss of generality, we limit our analysis to the case in which $ \partial f /\partial v_{\parallel} > 0$ at $v_{\parallel}>0$.

Plasma waves with a phase speed $v_{\mathrm{phase}}=\omega/k_{\parallel}$ can Landau-resonate with particles at $v_{\parallel}=v_{\mathrm{phase}}$, where $\omega$ is the wave frequency and $k_{\parallel}$ is the component of the wave vector parallel to the background magnetic field. If this Landau resonance occurs in the part of velocity space where $ \partial f /\partial v_{\parallel} > 0$,  the resonant interaction converts particle energy into wave energy, leading to wave growth \citep{baumjohann2012basic}. The inverse process, in which waves lose energy to the particles, corresponds to Landau damping. 

We approximate the Langmuir dispersion relation through the Bohm--Gross approximation \citep{chen2012introduction}
\begin{equation}\label{LWDR}
\omega^2 \approx \omega_{\mathrm{pe}}^2 + \frac{3}{2}k_{\parallel}^{2} v_{\mathrm{th,e}}^2,
\end{equation}
where $v_{\mathrm{th,e}} = \sqrt{2k_{\mathrm B} T_{\mathrm e}/m_{\mathrm e}}$ is the electron thermal speed, $k_{\mathrm B}$ is the Boltzmann constant, $T_{\mathrm e}$ is the electron temperature, and $m_{\mathrm e}$ is the electron mass. 

In the majority of solar wind intervals, strahl electrons are located very close to the core, thus not forming a positive velocity gradient in the VDF. Electron distributions, in the absence of energetic electron beams, are therefore unlikely to show a bump-on-tail configuration. 

Equation~(\ref{LWDR}) indicates that the second term $\propto v_{\mathrm{th,e}}^2$ plays an increasingly significant role as $k_\parallel$ increases. For $k_\parallel\rightarrow \infty$, the dispersion relation asymptotically approaches $v_{\mathrm{phase}}=\sqrt{3/2}v_{\mathrm{th,e}}$. Consequently, the bump-on-tail instability can only occur if $ \partial f /\partial v_{\parallel} > 0$ at $v_{\parallel}> v_\mathrm{thr} = \sqrt{3/2}v_{\mathrm{th,e}}$ \citep[see also][]{Pommois2017}. 

\subsection{The first adiabatic invariant -- magnetic moment $\mu$}\label{conditionsec}

When electrons traverse regions with inhomogeneous magnetic fields, their velocity distribution adjusts in response to these variations. The electrons conserve the magnetic moment $\mu$, which is the first adiabatic invariant, during this evolution when the electron is magnetized and changes in the magnetic field are slow compared to the electron gyration.
The magnetic moment $\mu$ is associated with the current created by the gyration of the electron \citep{ichimaru2018basic} and is given by
\begin{equation}\label{mu}
    \mu = \frac{m_\mathrm{e}v_\perp^2}{2B},
\end{equation}
where $B$ is the magnetic field strength.

When the magnetic field strength increases, conservation of $\mu$ demands that $v_{\perp}^2$ increases as well. The kinetic energy 
\begin{equation}
E=\frac{1}{2}m_{\mathrm e}(v_{\perp}^2+v_{\parallel}^2)
\end{equation}
remains conserved as long as no electric field is present. Therefore, $v_{\parallel}^2$ must decrease as $v_{\perp}^2$ increases. This effect leads to magnetic mirroring in the spatially increasing $B$-field when particles cross $v_\parallel = 0$. Localized depletion in $B$, such as MHs, are by definition bordered by regions with increasing $B$ on both sides, suggesting that mirroring occurs on both sides. Particles that mirror on both sides of an MH are trapped in this structure.

Particles with small $|v_{\perp}/v_{\parallel}|$ are not mirrored but instead move along the field lines. They enter and leave localized depletion in $B$. The loss cone defines the region in velocity space that is occupied by particles that leave localized depletion in $B$. Particles outside the loss cone form the trapped population.  
We define the loss-cone angle $\alpha$ through \citep{chen2012introduction}
\begin{equation}\label{trapangle}
    \sin^2 \alpha = \frac{B}{B_{\max}},
\end{equation}
where $B_\mathrm{max}$ is the maximum magnetic field strength of the structure. A particle is not mirrored if its pitch angle $\phi$ satisfies $\sin^2 \phi < \sin^2 \alpha$, where $\tan \phi = v_\perp/v_\parallel$.

\subsection{Breaking of $\mu$-conservation}

Formally, an electron must fulfill two conditions in order to conserve its magnetic moment when passing a inhomogeneous magnetic-field structure: (i) the particle gyro-radius must be smaller than the size of the magnetic structure, and (ii) the particle must be able to complete at least one gyration while crossing the structure in order to remain magnetized. 

Both conditions are fundamentally tied to the magnetic field geometry. To mathematically formulate these conditions, we first estimate the spatial size $R_{\mathrm c}$ of the field variation associated with the structure  as
\begin{equation}
    \frac{1}{R_{\mathrm c}} \sim \left|\frac{1}{B} \frac{\mathrm{d}{B}}{\mathrm{d}s}\right|,
\end{equation}
where $s$ is the spatial coordinate along the structure. We assume that the structure is in steady-state and propagates with the same speed as the proton bulk flow across our measurement point.
According to Taylor's hypothesis and the frozen-in theorem \citep{baumjohann2012basic}, we relate spatial derivatives to time derivatives in spacecraft measurements using the proton speed $U_{\mathrm p}$ as
\begin{equation}
    \frac{\mathrm{d}}{\mathrm{d}s} = \frac{1}{ U_{\mathrm p}} \frac{\mathrm{d}}{\mathrm{d}t}.
\end{equation}
Therefore,
\begin{equation} \label{rc}
    \frac{1}{R_{\mathrm c}} \sim \left|\frac{1}{B} \frac{\mathrm{d}{B}}{\mathrm d s}\right|\approx \left| \frac{1}{BU_{\mathrm p}} \frac{\mathrm{d}{B}}{\mathrm{d}t}\right|=\left|\frac{1}{U_{\mathrm p}} \frac{\mathrm{d}}{\mathrm{d}t} \ln B\right|.
\end{equation}

With this estimate for size of the structure, we express condition (i) as
\begin{equation}\label{con1}
     r_\mathrm{g} \lesssim R_\mathrm{c},
\end{equation}
where $r_\mathrm{g} = m_\mathrm{e} v_{\perp} c / e B$ is the electron gyro-radius, $c$ is the speed of light, and $e$ is the elementary charge. 
We rewrite condition (i) under the assumption that $U_{\mathrm p}>0$ as
\begin{equation}\label{con1simple}
    \sin \phi \lesssim \frac{U_{\mathrm{p}}}{\left|\displaystyle\frac{\mathrm{d} \ln{B}}{\mathrm{d} t}\right|} \frac{e B}{c} \sqrt{\frac{1}{2 m_\mathrm{e}E}}.
\end{equation}

We write condition (ii) as
\begin{equation}\label{con2}
    \tau \gtrsim  \frac{1}{\Omega_e};
\end{equation}
where $\tau$ is the crossing time of a given electron through the structure, and $\Omega_e = e B / m_{\mathrm{e}} c$ is the electron cyclotron frequency.
We estimate the time taken by a particle to cross the structure as
\begin{equation}
    \tau \sim \frac{R_{\mathrm{c}}}{v_{\parallel}} \sim \frac{U_{\mathrm{p}}}{\left|\displaystyle\frac{\mathrm{d} \ln{B}}{\mathrm{d}  t}\right|}\sqrt{\frac{m_\mathrm{e}}{2E}}\frac{1}{\cos \phi}.
\end{equation}
Therefore, we find that condition (ii) is fulfilled if
\begin{equation} \label{con2simple}
    \cos \phi \lesssim \frac{U_{\mathrm{p}}}{\left|\displaystyle\frac{\mathrm{d} \ln{B}}{\mathrm{d} t}\right|} \frac{e B}{c} \sqrt{\frac{1}{2 m_\mathrm{e}E}}.
\end{equation}

Both conditions, as expressed in Eqs.~\eqref{con1simple} and \eqref{con2simple}, define pitch-angle ranges at given energies $E$ in which the conditions are fulfilled, depending on the properties of magnetic field structure. The left-hand sides of both equations are  bounded between 0 and 1. The right-hand sides of Equations~ \eqref{con1simple} and \eqref{con2simple} are  greater than 1  in most cases in the solar wind, indicating that $\mu$ is conserved most of the time.  

When particles encounter a MH, one or both of these conditions can be violated.
In a deep MH, the magnetic field strength $B$ decreases significantly. In addition, if the change of the magnetic field occurs on a small spatial scale, $|\mathrm{d}\ln{B}/\textrm{d} t|$ is large, corresponding to a sharp MH. These two scenarios, deep MHs and sharp magnetic field gradients, may occur individually or simultaneously, potentially lowering the right-hand sides of Equations~\eqref{con1simple} and \eqref{con2simple} to values between 0 and $1$. When this occurs, the conservation of the magnetic moment in the structure is violated.

For constant $B$ and $\phi$, electrons at higher $E$ are more likely to break $\mu$-conservation according to these conditions. We define the maximum energy required to fulfill Equations~\eqref{con1simple} and \eqref{con2simple} as 
\begin{equation}
    E_\mathrm{crit} = \frac{m_{\mathrm e}}{2}\left( \frac{U_p}{\displaystyle\frac{\mathrm{d} \ln{B}}{\mathrm{d}t}}\right)^{2}\left(\frac{e B}{m_{\mathrm e}c}\right)^{2}.
\end{equation}
All electrons with $E \geq E_{\mathrm{crit}}$ have the potential to break $\mu$-conservation at a specific pitch angle.

When one or both conditions are not satisfied for electrons in a given part of velocity space, these electrons are no longer magnetized, and these electrons do not follow the adiabatic $\mu$-conservation anymore. We assume that these particles tend to remain in their original position in velocity space rather than being modified by the constraints imposed by the conservation of $\mu$. 

\subsection{Modification of VDF due to breaking of $\mu$-conservation}

\begin{figure}
    \centering
    \includegraphics[width=\linewidth]{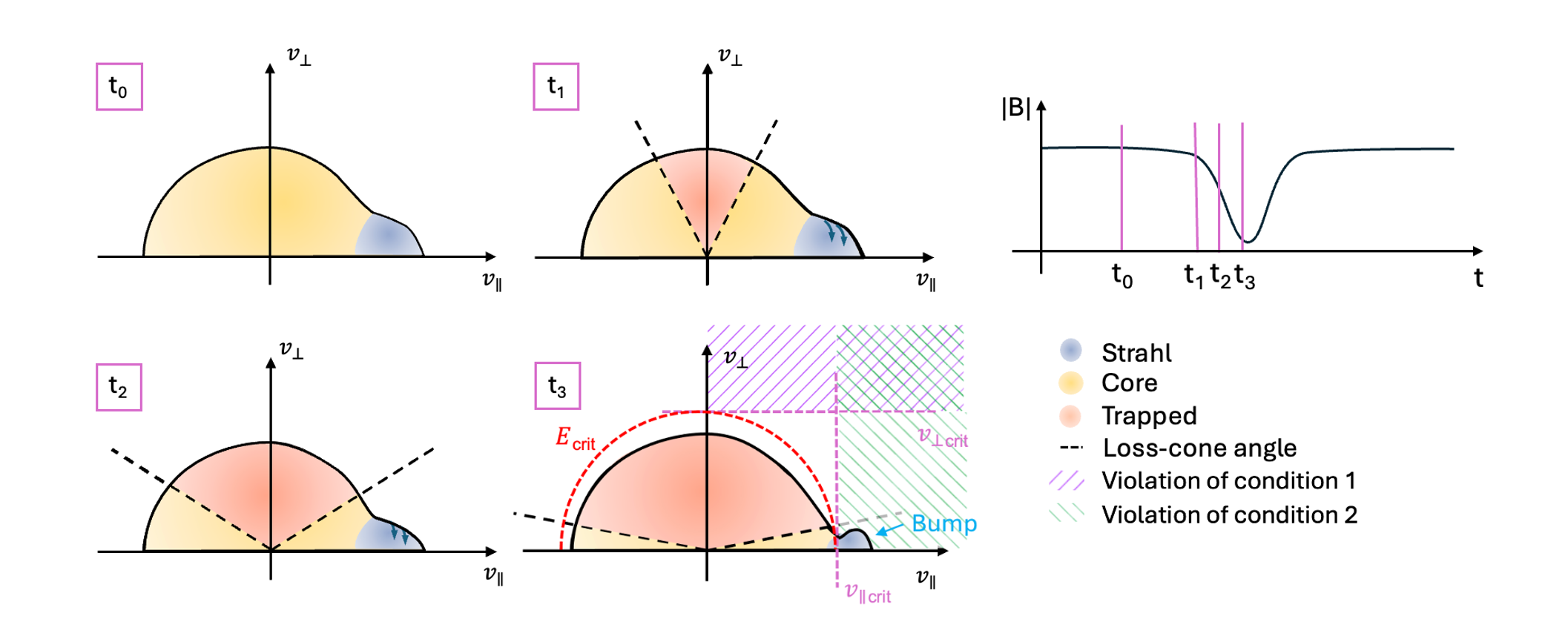}
    \caption{Illustration of changes in the electron VDF when passing through a MH. The variation in magnetic field strength as a function of time is shown on the right, with the times of interest indicated by pink color lines.
    Time $t_0$ corresponds to the time before entering the MH. Electrons with $v_{\parallel}>0$ stream towards the MH with a core and strahl configuration, while the distribution at $v_{\parallel}<0$ consists of core electrons that have already crossed the MH. For the subsequent times, the loss-cone angle $\alpha$ is represented by black dashed lines. The distribution functions at $t_1$, $t_2$, and $t_3$ illustrate the  VDFs at their respective locations within the MH.
    Near the minimum of $|B|$ at $t_3$, strahl electrons with velocities within the shaded area  violate their $\mu$-conservation. These electrons  have energy greater than the critical energy $E_{\mathrm{crit}}$ and form a bump in the form of a positive gradient in velocity space.}
    \label{cartoon}
\end{figure}

We now examine the effects of the breaking of the $\mu$-conservation on the overall electron VDF, with a particular focus on the strahl population. We illustrate our model in Figure~\ref{cartoon}. 

We build a scenario in which a spacecraft travels through a symmetric MH, as illustrated on the right panel in Figure \ref{cartoon}. At time $t_0$, we measure an electron distribution function for which  electrons with $v_{\parallel}>0$ move toward the MH and electrons with $v_{\parallel}<0$  have already traversed the structure. The strahl electrons in blue move towards the MH.

As electrons travel through the MH, they first encounter a reduction in the magnetic field strength. To conserve their magnetic moment $\mu$, the particles' perpendicular velocity decreases. In response, their parallel velocity increases to maintain the conservation of $E$. This change in magnetic field strength alters the velocity distribution function, causing the pitch-angle of electrons to decrease. This effect has no impact on the isotropic core electron population, as their $f$ depends on $v^2$ only. 
In contrast, strahl electrons are significantly influenced by the $\mu$-conservation. At time $t_1$, the strahl electrons occur at smaller pitch angles compared to their occurrence at $t_0$.

Electrons outside the loss cone are trapped within the MH. As our spacecraft moves to regions of decreasing magnetic field strength, the loss-cone angle decreases and the area of velocity space occupied by trapped electrons expands.
Electrons originating from outside the MHs are always within the loss cone, assuming they behave adiabatically, which allows them to enter and leave the structure. As they pass through the MH, these electrons experience a reduction in their pitch angles until they reach the bottom of the MH; however, they return to their initial velocity-space coordinates when leaving the MH on the other side.

At time $t_3$, we encounter the VDF at the minimum of $B$. At this point, the $\mu$-conservation is violated for the part of the electron VDF that does not satisfy Equations~\eqref{con1} and \eqref{con2}. 
By rearranging Equations~\eqref{con1simple} and \eqref{con2simple}, we express the conditions in terms of critical velocities:
\begin{equation} \label{vperpcrit}
    v_{\perp , \mathrm{crit}} = \sqrt{\frac{2 E_{\mathrm{crit}}}{m_{\mathrm{e}}}}\sin(90^{\circ}) = \sqrt{\frac{2 E_{\mathrm{crit}}}{m_{\mathrm{e}}}},
\end{equation}
and
\begin{equation}\label{vparacrit}
    v_{\parallel , \mathrm{crit}} = \sqrt{\frac{2 E_{\mathrm{crit}}}{m_{\mathrm{e}}}}\cos(0^{\circ}) = \sqrt{\frac{2 E_{\mathrm{crit}}}{m_{\mathrm{e}}}}.
\end{equation}

We indicate the velocity space at $v_{\parallel}\ge v_{\parallel , \mathrm{crit}}$ and $v_{\perp}\ge v_{\perp , \mathrm{crit}}$ as shaded areas in Figure~\ref{cartoon} for time $t_3$, indicating that any particles with velocities exceeding these critical values can violate the $\mu$-conservation. The adiabatic evolution of $f$ is restricted to the region where the conservation of $\mu$ is maintained.
As per our assumption strahl electrons that do not conserve $\mu$ anymore tend to appear at larger pitch angles than expected from $\mu$-conservation. If this deviation occurs near or above the transition from the core to the strahl part of the VDF, it can create a positive velocity gradient in the distribution function. When $\partial f/\partial v_\parallel>0 $ at velocities exceeding the bump-on-tail instability threshold $v_\mathrm{thr}$, the bump on the VDF can drive the Landau-resonant instability of LWs.

\section{Observations} \label{obsresult}

With the advent of the new space missions Solar Orbiter and Parker Solar Probe, MHs are now being studied in environments close to the Sun, where background magnetic field strengths and plasma conditions differ significantly from near-Earth space \citep{chen2021macro,yu2021characteristics,boldu2023langmuir}. The data from these missions allow us to study the kinetic properties of MHs in great detail. 

\subsection{Overview}

The dataset used for our research is collected from three instruments onboard Solar Orbiter: the Solar Wind Analyser (SWA), the Radio and Plasma Waves (RPW) instrument, and the Magnetometer (MAG). 
The Electron Analyser System (EAS) is part of the SWA suite and measures solar wind electrons at energies from a few eV to 5000\,eV, providing detailed three-dimensional VDFs; the Proton-Alpha Sensor (PAS) is also part of SWA and measures the VDFs of protons and $\alpha$-particles \citep{owen2020solar, owen2021high}. The RPW instrument captures variations in the electric and magnetic fields. It is capable of measuring in-situ waves with frequencies up to several hundred kHZ. RPW's Thermal Noise Receiver (TNR) delivers electromagnetic spectra from several kHz to 16\,MHz, and its Time Domain Sampler (TDS) serves as a medium-frequency receiver dedicated to capturing waveform data \citep{maksimovic2020solar,soucek2021solar}. The MAG instrument records vector magnetic field data at a sampling rate of 128 vectors/second \citep{horbury2020solar}. These instruments collectively provide the high-resolution measurements adopted for this investigation.

Based on our model presented in Section~\ref{conditionsec}, we identify two MH events in which the minimum magnetic field strength, as measured by the MAG instrument, decreases below 1\,nT. 

Figures~\ref{case1} and \ref{case2} summarize the relevant measurements. For both cases, panels (a) and (b) show the electron number density and the magnetic field components in RTN coordinates. Panel (f) shows a magnetic field feather plot to provide additional context of the local magnetic field structure. 

We plot the electron pitch-angle distribution from SWA/EAS focusing on strahl electrons in panel (c) for energies greater than 69\,eV. To estimate the size of the MH structure, we apply Equation~\eqref{rc}, using magnetic field measurements and proton speed measurements obtained from MAG and SWA/PAS.  We then evaluate Eq.~\eqref{con1simple} and \eqref{con2simple}.

In panel (e), we show the voltage power spectral density data and radio dynamic spectra from RPW-TNR. In these spectra, wave packages appear as localized enhancements.
The RPW-TDS captures the waveform of the observed waves as shown in the right panels of Figures~\ref{case1dr} and \ref{case2dr}, offering higher resolution electric field measurements. 

We show three instances of the two-dimensional electron VDF as a function of speed $v$ and pitch-angle $\phi$ in Figure~\ref{vdf1} for Case 1 and in Figure~\ref{vdf2} for Case 2. The first panel shows the VDF before the strahl electrons enter the MH. The second panel shows the VDF at the location of the minimum in the magnetic field strength. The third panel shows the VDF after the strahl electrons have left the MH.

\subsection{Case 1}

\begin{figure}
    \centering
    \includegraphics[width=0.9\linewidth]{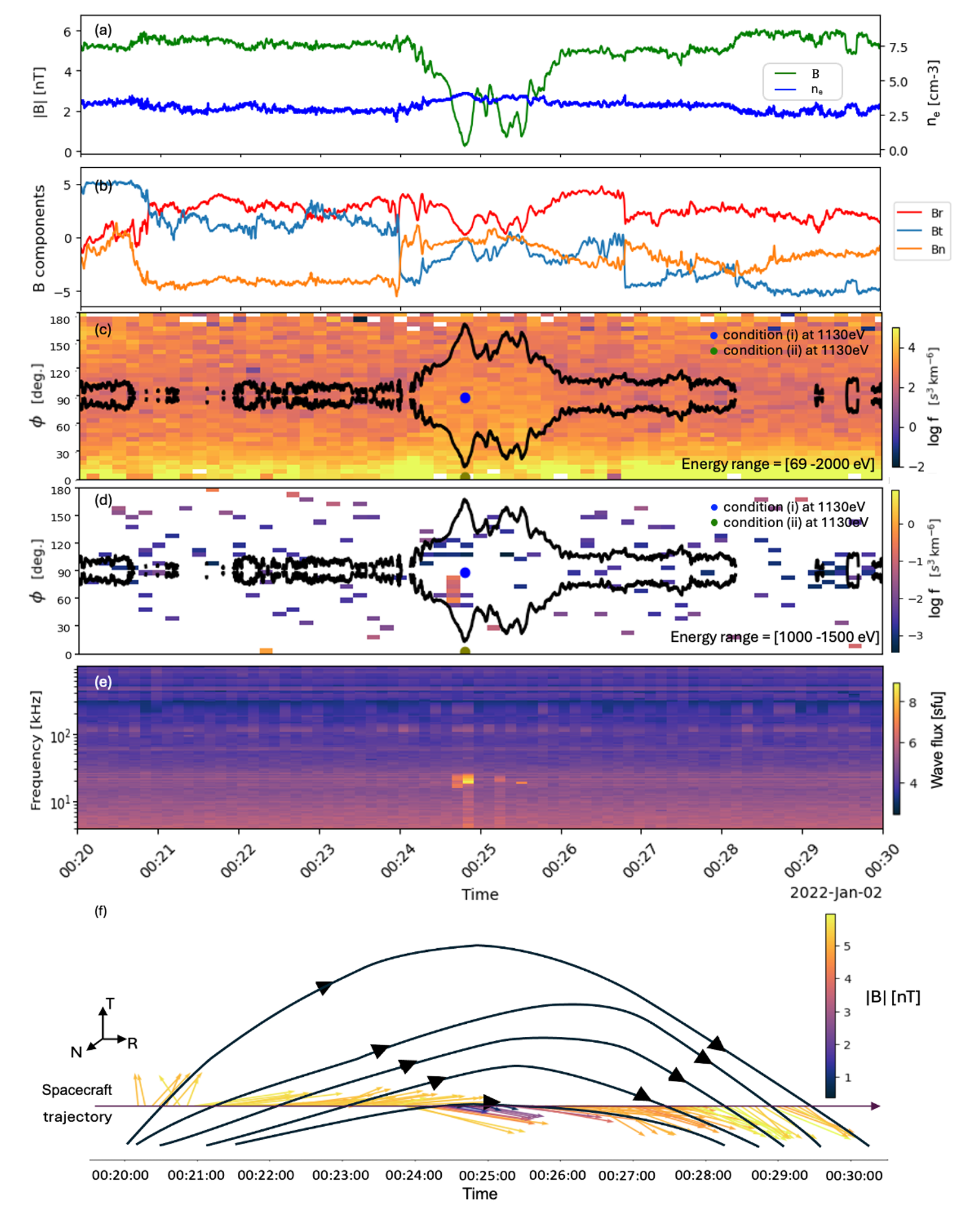}
    \caption{Solar Orbiter Observations of an MH on 2022 January 02 from 00:20:00 UT to 00:30:00 UT. The horizontal axis shows time in UT (hh:mm).
    (a) Magnetic field strength $B$ in green and electron number density $n_e$ in blue. 
    (b) Components of magnetic field in RTN coordinates, where $\mathrm{R}$ is the radial direction, $\mathrm{T}$ is the tangential direction, and $\mathrm{N}$ completes the right-handed triad. 
    (c) Electron pitch-angle distribution of electrons from 69\,eV to 2000\,eV  (d) Electron pitch-angle distribution from 1000\,eV to 1500\,eV. The blue and olive colored dots indicate where Equations~\eqref{con1simple} and  \eqref{con2simple} are violated for $E = E_{\mathrm{crit}}=1130\,\mathrm{eV}$.
    (e) Dynamic spectrum of wave flux versus time in spectral flux units (sfu).
    (f) Illustration of a possible magnetic field configuration based on a feather plot based on our in-situ magnetic field measurements. In this panel, time is displayed in hh:mm:ss format.}
    \label{case1}
\end{figure}

\begin{figure}
    \centering
    \includegraphics[width=\linewidth]{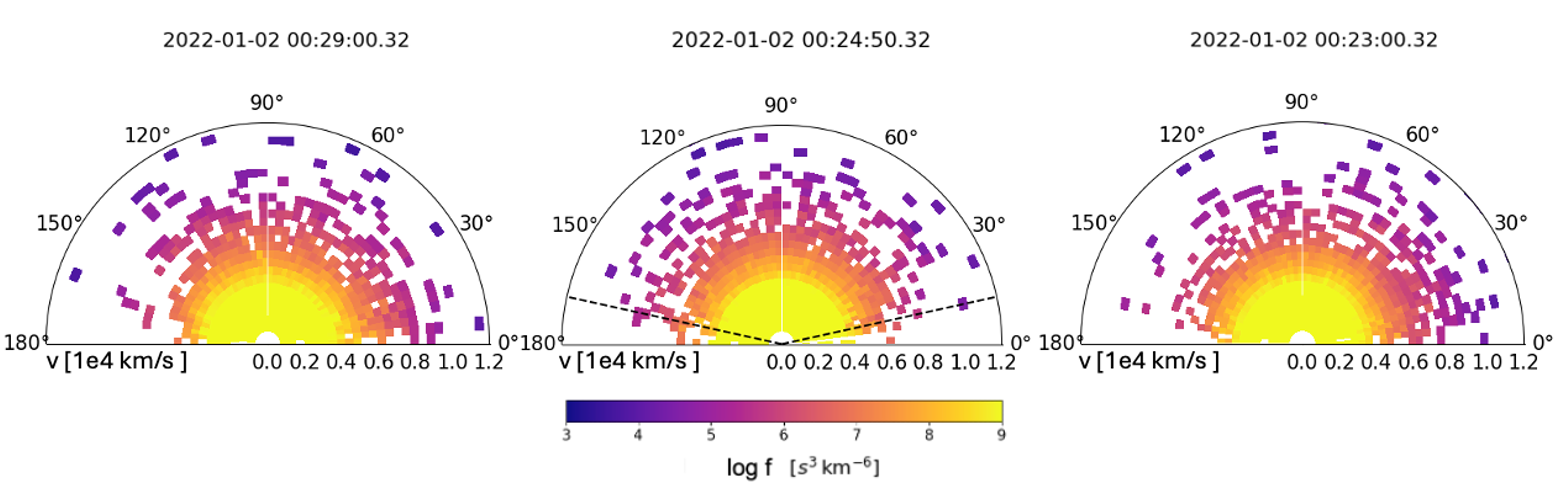}
    \caption{Electron velocity distribution functions at three characteristic phases for Case 1: (left) before the strahl population enters the MH, (center) at the time of the recorded minimum magnetic field strength with loss-cone angle indicated by black dashed lines, and (right) after the strahl population exits the MH. The corresponding times in UT (hh:mm:ss) are indicated above each panel. The angles indicate the pitch-angle $\phi$.}
    \label{vdf1}
\end{figure}

\begin{figure}
    \centering
    \includegraphics[width=\linewidth]{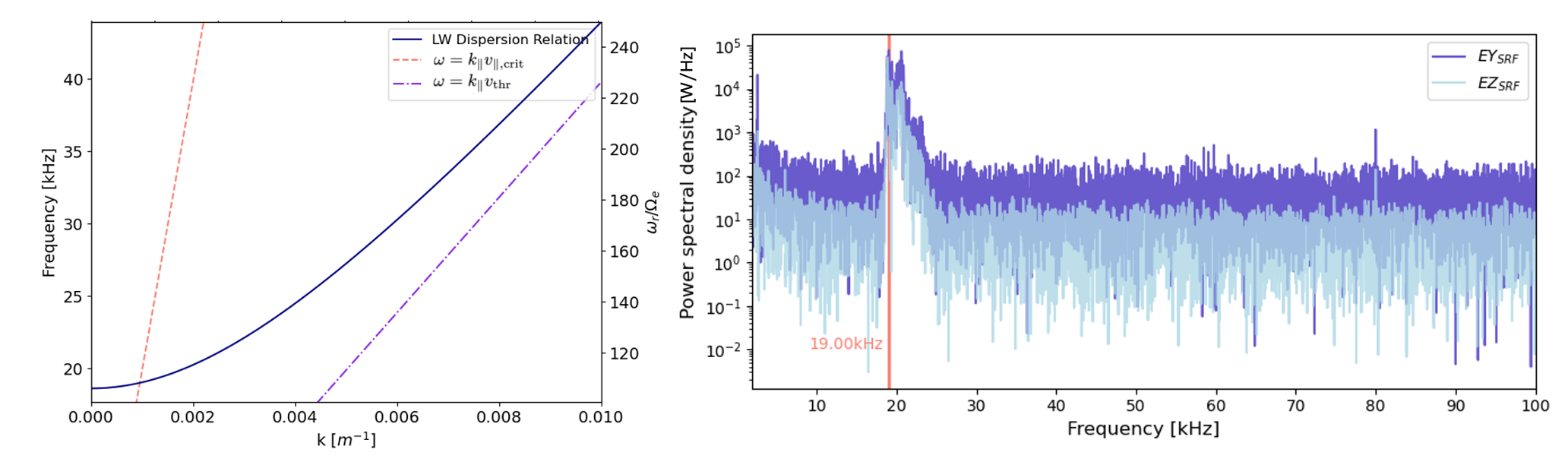}
    \caption{Left: LW dispersion relation from Equation~\eqref{LWDR}. We overplot the resonance condition $\omega=k_{\parallel}v_{\parallel,\mathrm{crit}}$ for $E_{\mathrm{crit}}=1130\,\mathrm{eV}$ in orange. The purple line shows $\omega=k_{\parallel}v_{\mathrm{thr}}$. Right: Triggered snapshot waveform from the RPW-TDS instrument, data captured at 00:24:52.79 UT on 2022 January 02.} $EY_{SRF}$ and $EZ_{SRF}$ represent the electric field components measured along the Y and Z axes in the spacecraft reference frame.
    \label{case1dr}
\end{figure}

As illustrated in Figure~\ref{case1}, the magnetic field strength drops from 5.723\,nT at 00:24:05 UT to 0.253\,nT at 00:24:49 UT. The black lines in the pitch-angle spectrum shown in panel (c) represent the loss-cone angles calculated with Equation~\eqref{trapangle}. We observe that the VDF values outside the loss cone (i.e., within the black lines) are greater than outside, indicating the presence of an enhanced trapped electron population. 

We calculate Equations~\eqref{con1simple} and \eqref{con2simple} throughout the time interval for this event, covering an energy range from 69\,eV to 2000\,eV. Within this energy range, electrons can only break our conditions for $\mu$-conservation  during the period of minimum magnetic field strength (occurring at 00:24:49 UT) with $E_{\mathrm{crit}}\approx 1130\,\mathrm{eV}$. The corresponding pitch-angle solutions from Equations~\eqref{con1simple} and \eqref{con2simple} are over-plotted in panels (c) and (d) as blue and olive data points.

In panel (d), we present the pitch-angle distribution for electrons in the energy range around $E_\mathrm{crit}$, spanning from 1000\,eV to 1500\,eV. We select this energy range around the calculated value of $E_\mathrm{crit}$ to account for uncertainties in our model predictions and to include more electron counts at high energies. We discuss the impact of finite particle counting statistics on our observational results in Section~\ref{limitation}.
This panel reveals a group of electrons with phase-space density significantly greater than the surrounding average, particularly at times near the minimum $B$. At the same time when Equations~\eqref{con1simple} and \eqref{con2simple} are not satisfied, RPW detects an enhancement in the electric field fluctuations at approximately 15\,kHz, as shown in panel (e).

The two-dimensional VDFs presented in Figure~\ref{vdf1} reveal the  evolution of the electron VDF. Before the strahl enters the MH (left) and after the strahl exits the MH (right), the VDFs exhibit similar distributions, both showing enhanced electron fluxes at small pitch angles around $\phi\gtrsim 0^{\circ}$. At the location of the recorded minimum magnetic field strength (center), we observe fewer electrons around $\phi\gtrsim 0^{\circ}$ and more electrons at higher energies and $\phi$. At the same time, the loss-cone angle reaches the range of pitch-angles occupied by strahl electrons.

Based on our model, we assume that the resonance velocity of LWs is near $v_{\parallel,\mathrm{crit}}$. By substituting $v_{\parallel , \mathrm{crit}}$ into Eq.~\eqref{LWDR}, we predict a LW frequency of approximately 19\,kHz at resonance. The resonance condition is fulfilled for those waves in Figure~\ref{case1dr} left for which the plot of the dispersion relation intersects the line marking $\omega=k_{\parallel}v_{\parallel , \mathrm{crit}}$.

We mark this predicted resonant frequency also in the panel on the right-hand side of Figure~\ref{case1dr} with a red vertical line. The right panel in Figure~\ref{case1dr} displays the power spectrum of the captured waveform, revealing the actual observed wave frequency. The spectrum shows a clear enhancement of electric-field fluctuations at and above the predicted resonant frequency.

\subsection{Case 2}

\begin{figure}
    \centering
    \includegraphics[width=0.9\linewidth]{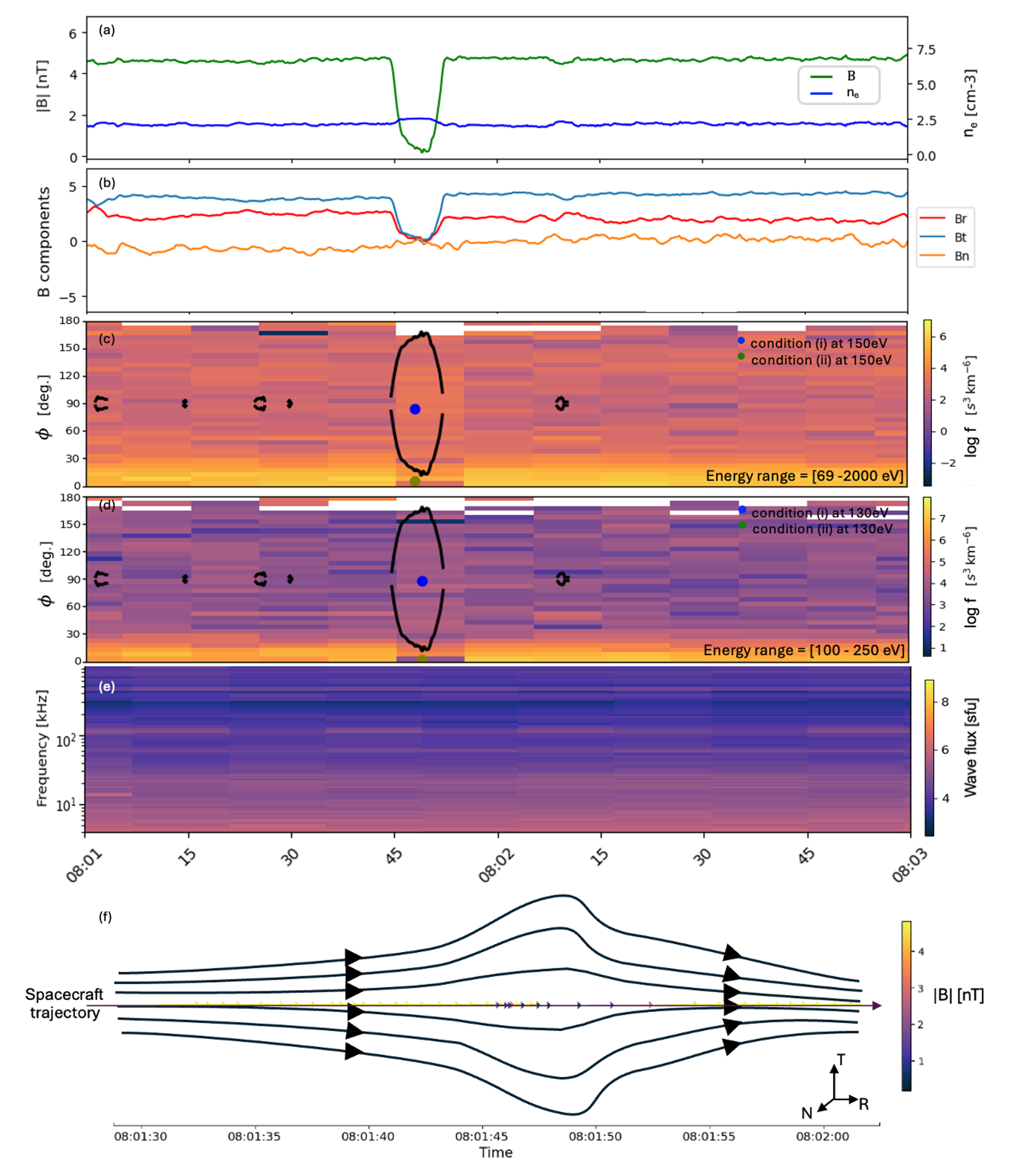}
    \caption{Solar Orbiter Observations of a MH on 2022 January 02 from 00:08:01 UT to 00:08:03 UT. The  horizontal axis shows time in UT (hh:mm).
    (a) Magnetic field strength $B$ in green and electron number density $n_e$ in blue.
    (b) Components of magnetic field in RTN coordinates, where $\mathrm{R}$ is the radial direction, $\mathrm{T}$ is the tangential direction, and $\mathrm{N}$ completes the right-handed triad. 
    (c) Electron pitch-angle distribution of electrons from 69\,eV to 2000\,eV. (d) Electron pitch-angle distribution from 100\,eV to 250\,eV. The blue and olive colored dots indicate where Equations~\eqref{con1simple} and  \eqref{con2simple} are violated for $E = E_{\mathrm{crit}}=150\,\mathrm{eV}$.
    (e) Dynamic spectrum of wave flux versus time in spectral flux units (sfu).
    (f) Illustration of a possible magnetic field configuration based on a feather plot based on our in-situ magnetic field measurements. In this panel, time is displayed in hh:mm:ss format.}
    \label{case2}
\end{figure}

\begin{figure}
    \centering
    \includegraphics[width=\linewidth]{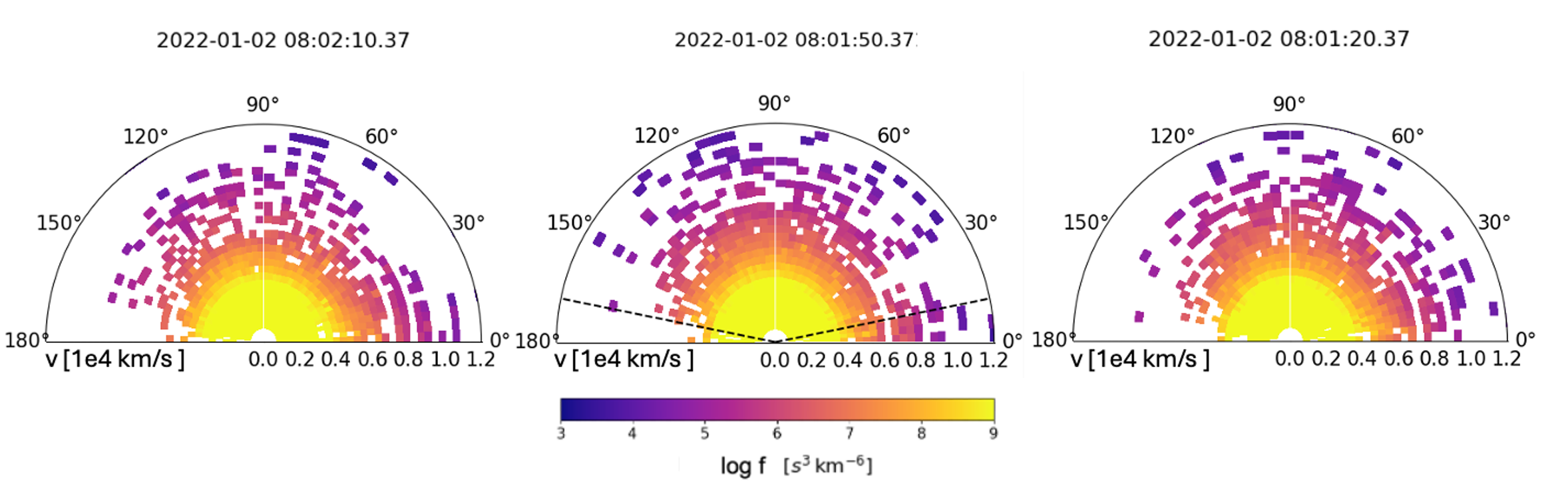}
    \caption{Electron velocity distribution functions at three characteristic phases for Case 2: (left) before the strahl population enters the MH, (center) at the time of the recorded minimum magnetic field strength with loss-cone angle indicated by black dashed lines, and (right) after the strahl population exits the MH. The corresponding times in UT (hh:mm:ss) are indicated above each panel. The angles indicate the pitch-angle $\phi$.}
    \label{vdf2}
\end{figure}

\begin{figure}
    \centering
    \includegraphics[width=\linewidth]{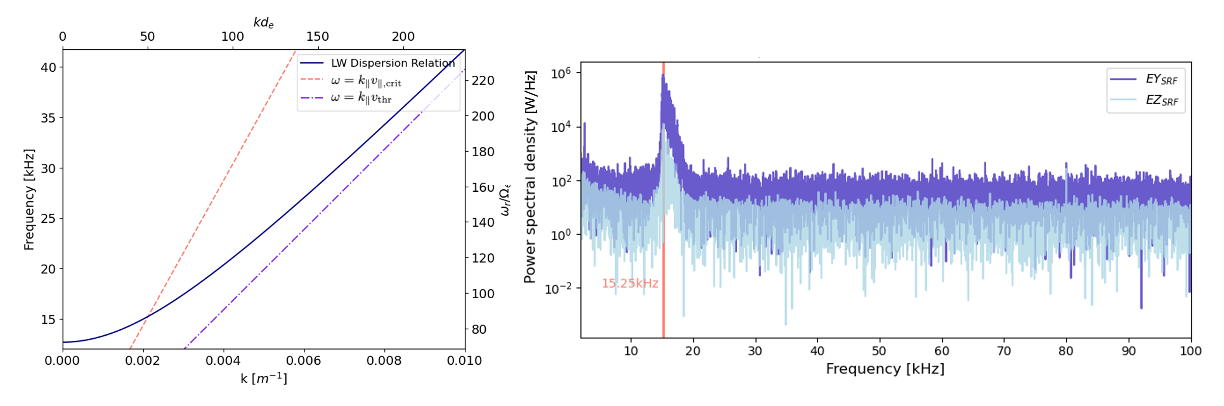}
    \caption{Left: LW dispersion relation from Equation~\eqref{LWDR}. We overplot the resonance condition $\omega=k_{\parallel}v_{\parallel,\mathrm{crit}}$ for $E_{\mathrm{crit}}=150\,\mathrm{eV}$ in orange. The purple line shows $\omega=k_{\parallel}v_{\mathrm{thr}}$. Right: Triggered snapshot waveform from the RPW-TDS instrument, data captured at 08:01:45.84 UT on 2022 January 02.} $EY_{SRF}$ and $EZ_{SRF}$ represent the electric field components measured along the Y and Z axes in the spacecraft reference frame.
    \label{case2dr}
\end{figure}

We study a second case of a MH associated with electric field fluctuations that are consistent with our model predictions. This MH has a shorter duration in the spacecraft frame.  The magnetic field variation decreases from 4.75\,nT at 08:01:43 UT. The field strength reaches its minimum value of 0.177\,nT at 08:01:49 UT. As shown in panel (f) of Figure~\ref{case2}, the consistently interpreted magnetic structure based on the feather plot of the magnetic field differs notably from that of Case 1. 

The decrease in $B$ occurs more rapidly compared to Case 1, resulting in a smaller $E_\mathrm{crit}\approx 150\,\mathrm{eV}$. In panel (d), we plot the electron pitch-angle distribution over the energy range of 100\,eV to 250\,eV. We observe that the electron VDF values at larger pitch angles are slightly greater than before and after the MH event. 

The dynamic spectrum in panel (e) does not show a clear enhancement during this time interval. However, the TDS triggered waveform snapshot successfully observe activity at frequencies around 15.7\,kHz as shown in Figure~\ref{case2dr}.

The two-dimensional VDFs in Figure~\ref{vdf2} show a similar trend to Case 1: enhanced higher-energy particle fluxes at larger pitch angles around the time of the recorded minimum in $B$ (center panel). At the same time, the loss-cone angle reaches the pitch-angle range occupied by strahl electrons. This observation is consistent with the electron dynamics shown in panel (d) of Figure~\ref{case2}.
 
For $E_{\mathrm{crit}} \approx 150 \, \mathrm{eV}$, the LW dispersion relation from Eq.~\eqref{LWDR} predicts a resonant frequency of approximately 15.7\,kHz. This predicted frequency is indicated by the orange color vertical line in the right panel of Figure~\ref{case2dr}, superimposed on the observed waveform snapshot.
The captured waveform exhibits a peak at 15.7\,kHz, which approximately aligns with our prediction based on the resonance condition and the dispersion relation for LWs.

\section{Discussion}

In this section, we provide our interpretation and further implications of our study. We discuss the limitations of our analysis and provide alternative scenarios.

\subsection{Interpretation and implications of our results}

Previous studies report LWs associated with magnetic depressions, with most cases occurring in MHs \citep{lin1996langmuir, lin1995observations, macdowall1996langmuir, boldu2023langmuir}. We present a model for the creation of LWs in MHs based on the violation of $\mu$-conservation for strahl electrons in the solar wind. Our model establishes threshold criteria in both pitch angle and energy for the violation of $\mu$-conservation. We present observations from Solar Orbiter that are consistent with the earlier observations of LWs near MHs and align with the predictions of our model. 

In the two cases we present, we find  enhancements in electrostatic fluctuations
during periods of low magnetic field strength. These enhancements occur at times at which our model predicts violation of $\mu$-conservation for strahl electrons. The observations are consistent with the proposed process of the creation of a bump-on-tail configuration which triggers LWs.

In Case 1, the MH has a longer duration in the spacecraft reference frame, allowing us to resolve  more details of the particle dynamics. We observe multiple LW events within the overall interval of Case 1. Most notably, a significant enhancement in LW activity coincides with the minimum of the recorded magnetic field strength, in agreement with our model prediction. Under quiet conditions, LWs usually arise as quasi-thermal noise caused by thermal electron motion, producing weak electrostatic fluctuations. The quasi-thermal noise forms peaks in the power spectrum near the plasma frequency \citep{le2009quasi}. The LW event at the time of the minimum of $B$, however, produces a clear enhancement of electrostatic fluctuations above the level of the quasi-thermal noise. While our model cannot predict all LW occurrences (as the physical mechanism may operate in regions beyond the spacecraft's trajectory), the clear enhancement of LWs at the field minimum offers a strong validation of our theoretical framework.

In Case 2, no clear LW signal is detected in the RPW-TDS radio spectrum, but the RPW-TNR captured a signature of LWs coinciding with the MH. This discrepancy may be due to the low time resolution of RPW-TNR. The TDS-tswf mode is designed to capture high-time-resolution data and is sensitive to transient phenomena. The MH in Case 2 is shorter in duration in the spacecraft frame and exhibits a sharper magnetic gradient compared to Case 1. The resulting increase in $R_{\mathrm c}$ leads to solutions to Equations~\eqref{con1} and \eqref{con2} at lower energy levels. The wider energy range in Case 2 allows more electrons to break $\mu$-conservation, potentially causing them to occupy a larger range of pitch angles. However, this does not necessarily help the development of a bump-on-tail configuration, as the increased population of electrons at lower energy levels may just raise the VDF values without creating a distinct positive gradient in $f$. Alternatively, a small bump may form, but it may not generate sufficiently large amplitudes, above the thermal noise, to be clearly detected by RPW-TNR.

Only particles above a specific energy can satisfy the condition for breaking of $\mu$-conservation in a MH. Most MHs have a minimum $B$ of about 0.1\,nT, so that electrons with energies exceeding $10^2\,\mathrm{eV}$ are more likely to meet the criteria. This energy threshold is significantly greater than the threshold energy required for the bump-on-tail instability, represented by $v_{\mathrm{thr}}$. Therefore, it is likely that strahl electrons in MHs above $E_{\mathrm{crit}}$, as long as they are present in sufficient number, trigger LWs effectively through our proposed mechanism. We note, however, that a locally anisotropic velocity distribution above $E_{\mathrm{crit}}$ is required for our model, as an isotropic distribution would not produce the necessary positive gradient in velocity space for the driving of the bump-on-tail instability.

According to our conditions for the violation of $\mu$-conservation, deep or sharp MHs are more likely to trigger LWs. However, not all deep or sharp MHs are observed in association with LWs. We attribute this observation to the number of electrons that  meet the criteria to break $\mu$-conservation. When there are too few resonant particles in the relevant part of velocity space, their energy transfer to LWs will result in fluctuations below the detection threshold.  
The calculated $E_\mathrm{crit}$ varies widely depending on the specific MH. A large $E_\mathrm{crit}$ is not always favorable for our mechanism, as the the number of particles participating in the violation of the magnetic moment is small.  However, if a bump-on-tail instability is triggered, the resonant velocity is likely to occur at energies close to $E_\mathrm{crit}$, as observed in Case 1 of our study. In contrast, a small $E_\mathrm{crit}$ value allows more electrons to violate $\mu$-conservation. 

Particles that break $\mu$-conservation can cross the trapping angle from the loss-cone into the trapped region of velocity space. Moreover, if LWs are excited, quasilinear diffusion reduces $v_\parallel$ of the resonant  electrons \citep{verscharen2022electron}, causing their phase-space distribution to migrate toward the thermal core population.
When these processes move particles from the loss-cone into the trapped region of velocity space, they serve as an additional source of the trapped population, feeding more particles into the structure.

\subsection{Limitations of our analysis}\label{limitation}

MHs are not static structures \citep{jiang2022whistler,helgesen1990dynamic}. Estimating their size and predicting their evolution based on  single-spacecraft measurements is very challenging. As a result, our calculation of $E_\mathrm{crit}$ based on the steady-state assumption and our estimate of $R_{\mathrm c}$ provides only an approximate result.
In fact, the properties of the magnetic field significantly influence our calculations. We estimate the size of the field depletion associated with the MH using the proton speed $U_{\mathrm p}$, and the change in the magnetic field strength is measured in the spacecraft reference frame. Both factors affect the determination of $R_{\mathrm c}$. 
A precise calculation of $R_{\mathrm c}$ would require detailed knowledge of the MH geometry, which is unavailable from single-spacecraft measurements. In this vein, we interpret our formulation of condition (i) and condition (ii) as approximate indicators for potential breaking of $\mu$-conservation rather than a rigid prediction of this process.

Higher-resolution data would significantly enhance our understanding of MHs. Higher-energy electrons are more likely to violate $\mu$-conservation in a MH, but only few electrons are detected in this energy range.
The measurement of these particles can be obscured by instrumental noise, making them difficult to resolve. 
This effect introduces uncertainty in the observations due to finite counting statistics \citep{nicolaou2024artificial}. Furthermore, the very short timescales associated with the quasilinear relaxation of the VDF through instabilities pose a challenge for the capturing of transient changes in the VDF with the current energy and time resolution of the available instruments. For instance, electron instabilities occur on timescales of order tens to hundreds of electron gyro-periods, corresponding to an approximate magnitude of $10^{-2}\sim 10^{-1}$\,s under our observed conditions \citep{boyd2003physics}. The Solar Orbiter SWA/EAS instrument records electron pitch-angle distributions at a cadence of 10\,s. Consequently, we do not expect to observe a distinct bump feature in Figures~\ref{vdf1} and \ref{vdf2}.
Due to this timescale separation, we cannot resolve the instability's growth phase directly. However, our observations provide sufficient indirect evidence for the occurrence of the proposed instability in the form of the detection of the unstable waves.
In fact, the pitch-angle distribution shown in Case 2 does not provide further insight into the electron behavior at the energy levels expected to trigger LWs. The spacecraft encounters a very small MH or merely crossed the boundary of a larger MH. Panel (f) in Figure~\ref{case2} provides a possible structure for this event, but the real structure may differ from our proposed representation. 

Beyond the particle measurements, the wave detection is also constrained by instrumental limitations. The LW intensity may fall below the detection threshold of the spacecraft instrumentation, rendering the waves unobservable despite their presence, as suggested by Case 2.

The specific trajectory with which the spacecraft crosses a MH is coincidental.
Multi-spacecraft missions would be highly beneficial for studies like ours by enabling the determination of both the wavenumber of the observed waves and the three-dimensional morphology of the observed MHs. Resolving the wave-vector of the unstable waves and the overall field geometry would help us confirm the resonance condition and provide deeper insights into the underlying physics \citep{wang2021first,huang2017statistical,sun2012cluster}.

\subsection{Alternative scenarios}

MHs are complex structures in which multiple parameters of the ambient plasma change simultaneously. While we propose and test one possible mechanism that triggers LWs in MHs, different mechanisms may come into play depending on the nature of the MHs. For instance, density gradients can influence the electron behavior and trigger LWs \citep{pechhacker2014three}. In the majority of MHs, observations indicate an increase in number density as the magnetic field strength decreases to maintain  pressure balance. However, the simulations explore density gradient that significantly exceed those observed in our cases. It would be worthwhile to adapt the simulations accordingly and to investigate the joint effects of field and density modulations on  the electron VDF in realistic MHs.

Previous studies establish mechanisms for the generation of whistler waves in MHs \citep{ahmadi2018generation,jiang2022whistler,xu2025excitation}; however, we do not observe whistler waves in our presented cases. These previous analyses suggest that whistler waves are generated by time-evolving MHs, in which electron trapping, the generation of temperature anisotropy, or the formation of electron beams play key roles. In our cases, however, steady-state MHs create LWs through instability when strahl electrons encounter the MH structure. Therefore, we consider our cases complementary to the whistler-wave-emitting cases.

Other processes, such as betatron cooling \citep{guo2021betatron,jiang2022whistler} and magnetic pumping \citep{lichko2020magnetic}, are reported in magnetospheric MHs. In our study, we do not observe such processes, which may be attributed to the distinct nature of quasi-stable MHs in the solar wind compared to those in the magnetosphere. It remains unclear whether these observed differences arise from the intrinsic properties or the size of the MHs under consideration, the properties of the strahl in the solar wind, or whether they are due to limitations in the instrument resolution in our observations.

\section{Conclusions}

We propose a model to explain the excitation of LWs in solar-wind MHs. Our model accounts for kinetic effects arising from the violation of the conservation of the magnetic moment that occurs when electrons from an asymmetric velocity distribution interact with localized magnetic field depressions.

Our model predicts conditions under which  some of the suprathermal electrons break $\mu$-conservation in MHs when these MHs are sufficiently deep. By making reasonable assumptions about the size of the field structure, we derive the critical energy $E_{\mathrm{crit}}$ above which electrons break $\mu$-conservation for given pitch angles.
Under certain conditions, these electrons  create localized beams in the electron distribution that drive LWs through the bump-on-tail instability within the MHs.

We present observations from the Solar Orbiter spacecraft that are consistent with our model. They show LWs  in regions for which our model predicts their occurrence.  The observed frequencies of the enhancement in the electric field power spectrum agree with the predicted frequency from our model based on the analytical dispersion relation and the resonance velocity of the beam electrons created by violation of the $\mu$-conservation. 

Although current spacecraft missions are limited in their capability to provide additional validation for our model, the cases shown here offer compelling evidence for our proposed mechanism for LW-emitting MHs. While our study focuses on MHs in the solar wind, MHs in planetary magnetospheres and other plasma environments may show similar behavior. However, the requirement of the presence of an anisotropic suprathermal feature like the strahl makes solar-wind MHs more likely to undergo our proposed mechanism. The violation of $\mu$-conservation in such environments could potentially trigger additional plasma processes beyond LW excitation due to the modification of the VDF.

Our work demonstrates that inhomogeneous field structures such as MHs play an important role for the transport of suprathermal electrons in collisionless plasmas. 
By establishing links between these structures and the particle behavior, our findings advance the understanding of energy conversion mechanisms. Furthermore, they highlight how magnetic structures critically define both particle distributions and wave activity in the plasma through the coupling of ion-scale and electron-scale kinetic processes. 
As extensions of our work, we propose statistical analyses based on our predictions and multi-point measurements with multi-spacecraft missions. This future work will advance our understanding of the excitation of LWs in solar-wind MHs and similar multi-scale couplings associated with plasma inhomogeneities. \\

\noindent 
This work was supported by STFC Consolidated Grant ST/W001004/1 and UKSA grant ST/X002012/1. This work was supported by the Royal Society (UK) and the Consiglio Nazionale delle Ricerche (Italy) through the International Exchanges Cost Share scheme/Joint Bilateral Agreement project ``Multi-scale electrostatic energisation of plasmas: comparison of collective processes in laboratory and space'' (award numbers IEC$\backslash{}$R2$\backslash{}$222050 and SAC.AD002.043.021). O.P. acknowledges the project
``2022KL38BK -- The ULtimate fate of TuRbulence from space to laboratory
plAsmas (ULTRA)'' (Master CUP B53D23004850006) by the Italian Ministry
of University and Research, funded under the National Recovery and Resilience
Plan (NRRP), Mission 4 -- Component C2 -- Investment 1.1, ``Fondo per il
Programma Nazionale di Ricerca e Progetti di Rilevante Interesse Nazionale
(PRIN 2022)'' (PE9) by the European Union – NextGenerationEU. W.J. is supported by the National Natural Science Foundation of China Grant No.~42404177, the NSSC Youth Grant, and the Specialized Research Fund for State Key Laboratories of China.

%

\vspace{5mm}








\bibliography{sample631}{}

\begin{thebibliography}{}
\expandafter\ifx\csname natexlab\endcsname\relax\def\natexlab#1{#1}\fi
\providecommand{\url}[1]{\href{#1}{#1}}
\providecommand{\dodoi}[1]{doi:~\href{http://doi.org/#1}{\nolinkurl{#1}}}
\providecommand{\doeprint}[1]{\href{http://ascl.net/#1}{\nolinkurl{http://ascl.net/#1}}}
\providecommand{\doarXiv}[1]{\href{https://arxiv.org/abs/#1}{\nolinkurl{https://arxiv.org/abs/#1}}}

\bibitem[{Ahmadi {et~al.}(2018)Ahmadi, Wilder, Ergun, Argall, Usanova, Breuillard, Malaspina, Paulson, Germaschewski, Eriksson, {et~al.}}]{ahmadi2018generation}
Ahmadi, N., Wilder, F.~D., Ergun, R., {et~al.} 2018, Journal of Geophysical Research: Space Physics, 123, 6383

\bibitem[{{Arr{\`o}} {et~al.}(2024){Arr{\`o}}, {Califano}, {Pucci}, {Karlsson}, \& {Li}}]{arro2024largescale}
{Arr{\`o}}, G., {Califano}, F., {Pucci}, F., {Karlsson}, T., \& {Li}, H. 2024, \apjl, 970, L6, \dodoi{10.3847/2041-8213/ad61da}

\bibitem[{{Arr{\`o}} {et~al.}(2023){Arr{\`o}}, {Pucci}, {Califano}, {Innocenti}, \& {Lapenta}}]{arro2023generation}
{Arr{\`o}}, G., {Pucci}, F., {Califano}, F., {Innocenti}, M.~E., \& {Lapenta}, G. 2023, \apj, 958, 11, \dodoi{10.3847/1538-4357/acf12e}

\bibitem[{Baumjohann \& Treumann(2012)}]{baumjohann2012basic}
Baumjohann, W., \& Treumann, R.~A. 2012, Basic space plasma physics (World Scientific)

\bibitem[{Bold{\'u} {et~al.}(2023)Bold{\'u}, Graham, Morooka, Andr{\'e}, Khotyaintsev, Karlsson, Sou{\v{c}}ek, P{\'\i}{\v{s}}a, \& Maksimovic}]{boldu2023langmuir}
Bold{\'u}, J., Graham, D., Morooka, M., {et~al.} 2023, Astronomy \& Astrophysics, 674, A220

\bibitem[{Boyd \& Sanderson(2003)}]{boyd2003physics}
Boyd, T. J.~M., \& Sanderson, J.~J. 2003, The physics of plasmas (Cambridge university press)

\bibitem[{Burlaga \& Lemaire(1978)}]{burlaga1978interplanetary}
Burlaga, L., \& Lemaire, J. 1978, Journal of Geophysical Research: Space Physics, 83, 5157

\bibitem[{Cairns(1986)}]{cairns1986source}
Cairns, I.~H. 1986, Publications of the Astronomical Society of Australia, 6, 444

\bibitem[{Chen {et~al.}(2021)Chen, Liu, \& Hu}]{chen2021macro}
Chen, C., Liu, Y.~D., \& Hu, H. 2021, The Astrophysical Journal, 921, 15

\bibitem[{Chen(2012)}]{chen2012introduction}
Chen, F.~F. 2012, Introduction to plasma physics (Springer Science \& Business Media)

\bibitem[{Feldman {et~al.}(1975)Feldman, Asbridge, Bame, Montgomery, \& Gary}]{feldman1975solar}
Feldman, W., Asbridge, J., Bame, S., Montgomery, M., \& Gary, S. 1975, Journal of Geophysical Research, 80, 4181

\bibitem[{Ge {et~al.}(2011)Ge, McFadden, Raeder, Angelopoulos, Larson, \& Constantinescu}]{ge2011case}
Ge, Y., McFadden, J., Raeder, J., {et~al.} 2011, Journal of Geophysical Research: Space Physics, 116

\bibitem[{Guo {et~al.}(2021)Guo, Fu, Cao, Fan, Yao, Liu, Chen, Wang, Liu, Xu, {et~al.}}]{guo2021betatron}
Guo, Z., Fu, H., Cao, J., {et~al.} 2021, Geophysical Research Letters, 48, e2021GL093826

\bibitem[{Helgesen {et~al.}(1990)Helgesen, Pieranski, \& Skjeltorp}]{helgesen1990dynamic}
Helgesen, G., Pieranski, P., \& Skjeltorp, A. 1990, Physical Review A, 42, 7271

\bibitem[{Herr(2016)}]{herr2016introduction}
Herr, W. 2016, arXiv preprint arXiv:1601.05227

\bibitem[{Horaites {et~al.}(2018)Horaites, Astfalk, Boldyrev, \& Jenko}]{horaites2018stability}
Horaites, K., Astfalk, P., Boldyrev, S., \& Jenko, F. 2018, Monthly Notices of the Royal Astronomical Society, 480, 1499

\bibitem[{Horbury {et~al.}(2020)Horbury, O’brien, Blazquez, Bendyk, Brown, Hudson, Evans, Oddy, Carr, Beek, {et~al.}}]{horbury2020solar}
Horbury, T., O’brien, H., Blazquez, I.~C., {et~al.} 2020, Astronomy \& Astrophysics, 642, A9

\bibitem[{Huang {et~al.}(2017)Huang, Du, Sahraoui, Yuan, He, Zhao, Le~Contel, Breuillard, Wang, Yu, {et~al.}}]{huang2017statistical}
Huang, S., Du, J., Sahraoui, F., {et~al.} 2017, Journal of Geophysical Research: Space Physics, 122, 8577

\bibitem[{Huang {et~al.}(2020)Huang, Xu, He, Jiang, Yuan, Deng, Wei, Zhang, \& Zhang}]{huang2020excitation}
Huang, S., Xu, S., He, L., {et~al.} 2020, Geophysical Research Letters, 47, e2020GL087515

\bibitem[{Ichimaru(2018)}]{ichimaru2018basic}
Ichimaru, S. 2018, Basic principles of plasma physics: a statistical approach (CRC Press)

\bibitem[{Jiang {et~al.}(2024)Jiang, Verscharen, Jeong, Li, Klein, Owen, \& Wang}]{jiang2024}
Jiang, W., Verscharen, D., Jeong, S.-Y., {et~al.} 2024, The Astrophysical Journal, 960, 30, \dodoi{10.3847/1538-4357/ad0df8}

\bibitem[{Jiang {et~al.}(2022)Jiang, Verscharen, Li, Wang, \& Klein}]{jiang2022whistler}
Jiang, W., Verscharen, D., Li, H., Wang, C., \& Klein, K.~G. 2022, The Astrophysical Journal, 935, 169

\bibitem[{Lapenta \& Bettarini(2011)}]{lapenta2011self}
Lapenta, G., \& Bettarini, L. 2011, Geophysical Research Letters, 38

\bibitem[{Le~Chat {et~al.}(2009)Le~Chat, Issautier, Meyer-Vernet, Zouganelis, Maksimovic, \& Moncuquet}]{le2009quasi}
Le~Chat, G., Issautier, K., Meyer-Vernet, N., {et~al.} 2009, Physics of Plasmas, 16

\bibitem[{Lichko \& Egedal(2020)}]{lichko2020magnetic}
Lichko, E., \& Egedal, J. 2020, Nature Communications, 11, 2942

\bibitem[{Lin {et~al.}(1995)Lin, Kellogg, MacDowall, Balogh, Forsyth, Phillips, Buttighoffer, \& Pick}]{lin1995observations}
Lin, N., Kellogg, P., MacDowall, R., {et~al.} 1995, Geophysical Research Letters, 22, 3417

\bibitem[{Lin {et~al.}(1996)Lin, Kellogg, MacDowall, Tsurutani, \& Ho}]{lin1996langmuir}
Lin, N., Kellogg, P.~J., MacDowall, R., Tsurutani, B., \& Ho, C. 1996, Astronomy and Astrophysics, v. 316, p. 425-429, 316, 425

\bibitem[{Lucek {et~al.}(1999)Lucek, Dunlop, Balogh, Cargill, Baumjohann, Georgescu, Haerendel, \& Fornacon}]{lucek1999mirror}
Lucek, E., Dunlop, M., Balogh, A., {et~al.} 1999, Geophysical research letters, 26, 2159

\bibitem[{MacDowall {et~al.}(1996)MacDowall, Lin, Kellogg, Balogh, Forsyth, \& Neugebauer}]{macdowall1996langmuir}
MacDowall, R., Lin, N., Kellogg, P., {et~al.} 1996, AIP Conference Proceedings, 382, 301

\bibitem[{Maksimovic {et~al.}(2020)Maksimovic, Bale, Chust, Khotyaintsev, Krasnoselskikh, Kretzschmar, Plettemeier, Rucker, Sou{\v{c}}ek, Steller, {et~al.}}]{maksimovic2020solar}
Maksimovic, M., Bale, S., Chust, T., {et~al.} 2020, Astronomy \& Astrophysics, 642, A12

\bibitem[{Nicolaou {et~al.}(2024)Nicolaou, Livadiotis, \& Ioannou}]{nicolaou2024artificial}
Nicolaou, G., Livadiotis, G., \& Ioannou, C. 2024, The Astrophysical Journal, 977, 168

\bibitem[{Owen {et~al.}(2020)Owen, Bruno, Livi, Louarn, Al~Janabi, Allegrini, Amoros, Baruah, Barthe, Berthomier, {et~al.}}]{owen2020solar}
Owen, C., Bruno, R., Livi, S., {et~al.} 2020, Astronomy \& Astrophysics, 642, A16

\bibitem[{Owen {et~al.}(2021)Owen, Kataria, Ber{\v{c}}i{\v{c}}, Horbury, Berthomier, Verscharen, Bruno, Livi, Louarn, Anekallu, {et~al.}}]{owen2021high}
Owen, C., Kataria, D., Ber{\v{c}}i{\v{c}}, L., {et~al.} 2021, Astronomy \& Astrophysics, 656, L9

\bibitem[{Pechhacker \& Tsiklauri(2014)}]{pechhacker2014three}
Pechhacker, R., \& Tsiklauri, D. 2014, Physics of Plasmas, 21

\bibitem[{{Pommois} {et~al.}(2017){Pommois}, {Valentini}, {Pezzi}, \& {Veltri}}]{Pommois2017}
{Pommois}, K., {Valentini}, F., {Pezzi}, O., \& {Veltri}, P. 2017, Physics of Plasmas, 24, 012105, \dodoi{10.1063/1.4973829}

\bibitem[{Pulupa {et~al.}(2020)Pulupa, Bale, Badman, Bonnell, Case, de~Wit, Goetz, Harvey, Hegedus, Kasper, {et~al.}}]{pulupa2020statistics}
Pulupa, M., Bale, S.~D., Badman, S.~T., {et~al.} 2020, The Astrophysical Journal Supplement Series, 246, 49

\bibitem[{Robinson \& Benz(2000)}]{robinson2000bidirectional}
Robinson, P., \& Benz, A. 2000, Solar Physics, 194, 345

\bibitem[{Russell {et~al.}(1987)Russell, Riedler, Schwingenschuh, \& Yeroshenko}]{russell1987mirror}
Russell, C., Riedler, W., Schwingenschuh, K., \& Yeroshenko, Y. 1987, Geophysical research letters, 14, 644

\bibitem[{Schroeder {et~al.}(2021)Schroeder, Boldyrev, \& Astfalk}]{schroeder2021stability}
Schroeder, J.~M., Boldyrev, S., \& Astfalk, P. 2021, Monthly Notices of the Royal Astronomical Society, 507, 1329

\bibitem[{Shi {et~al.}(2009)Shi, Pu, Soucek, Zong, Fu, Xie, Chen, Zhang, Li, Xia, {et~al.}}]{shi2009spatial}
Shi, Q., Pu, Z., Soucek, J., {et~al.} 2009, Journal of Geophysical Research: Space Physics, 114

\bibitem[{Soucek {et~al.}(2021)Soucek, P{\'\i}{\v{s}}a, Kolmasova, Uhlir, Lan, Santol{\'\i}k, Krupar, Kruparova, Ba{\v{s}}e, Maksimovic, {et~al.}}]{soucek2021solar}
Soucek, J., P{\'\i}{\v{s}}a, D., Kolmasova, I., {et~al.} 2021, Astronomy \& Astrophysics, 656, A26

\bibitem[{Sperveslage {et~al.}(2000)Sperveslage, Neubauer, Baumg{\"a}rtel, \& Ness}]{sperveslage2000magnetic}
Sperveslage, K., Neubauer, F., Baumg{\"a}rtel, K., \& Ness, N. 2000, Nonlinear Processes in Geophysics, 7, 191

\bibitem[{{Sun} {et~al.}(2012){Sun}, {Shi}, {Fu}, {Pu}, {Dunlop}, {Walsh}, {Zong}, {Xiao}, {Tang}, {Reme}, {Carr}, {Lucek}, \& {Fazakerley}}]{sun2012cluster}
{Sun}, W.~J., {Shi}, Q.~Q., {Fu}, S.~Y., {et~al.} 2012, Annales Geophysicae, 30, 583, \dodoi{10.5194/angeo-30-583-2012}

\bibitem[{Tsurutani {et~al.}(1982)Tsurutani, Smith, Anderson, Ogilvie, Scudder, Baker, \& Bame}]{tsurutani1982lion}
Tsurutani, B., Smith, E., Anderson, R., {et~al.} 1982, Journal of Geophysical Research: Space Physics, 87, 6060

\bibitem[{Turner {et~al.}(1977)Turner, Burlaga, Ness, \& Lemaire}]{turner1977magnetic}
Turner, J., Burlaga, L., Ness, N., \& Lemaire, J. 1977, Journal of Geophysical Research, 82, 1921

\bibitem[{Verscharen {et~al.}(2019{\natexlab{a}})Verscharen, Chandran, Jeong, Salem, Pulupa, \& Bale}]{verscharen2019self}
Verscharen, D., Chandran, B.~D., Jeong, S.-Y., {et~al.} 2019{\natexlab{a}}, The Astrophysical Journal, 886, 136

\bibitem[{Verscharen {et~al.}(2019{\natexlab{b}})Verscharen, Klein, \& Maruca}]{verscharen2019multi}
Verscharen, D., Klein, K.~G., \& Maruca, B.~A. 2019{\natexlab{b}}, Living Reviews in Solar Physics, 16, 5

\bibitem[{Verscharen {et~al.}(2022)Verscharen, Chandran, Boella, Halekas, Innocenti, Jagarlamudi, Micera, Pierrard, {\v{S}}tver{\'a}k, Vasko, {et~al.}}]{verscharen2022electron}
Verscharen, D., Chandran, B.~D., Boella, E., {et~al.} 2022, Frontiers in astronomy and space sciences, 9, 951628

\bibitem[{Wang {et~al.}(2021)Wang, Volwerk, Wu, Hao, Xiao, Wang, Liu, Chen, \& Zhang}]{wang2021first}
Wang, G., Volwerk, M., Wu, M., {et~al.} 2021, The Astronomical Journal, 161, 110

\bibitem[{Winterhalter {et~al.}(1994)Winterhalter, Neugebauer, Goldstein, Smith, Bame, \& Balogh}]{winterhalter1994ulysses}
Winterhalter, D., Neugebauer, M., Goldstein, B.~E., {et~al.} 1994, Journal of Geophysical Research: Space Physics, 99, 23371

\bibitem[{Xiao {et~al.}(2014)Xiao, Shi, Tian, Sun, Zhang, Shen, Shang, \& Du}]{xiao2014plasma}
Xiao, T., Shi, Q., Tian, A., {et~al.} 2014, Coronal magnetometry, 553

\bibitem[{Xu {et~al.}(2025)Xu, Wang, Fu, Fu, Zhang, Yu, Guo, \& Cao}]{xu2025excitation}
Xu, Z., Wang, Z., Fu, H., {et~al.} 2025, Journal of Geophysical Research: Space Physics, 130, e2024JA033524

\bibitem[{Yao {et~al.}(2019)Yao, Shi, Yao, Li, Yue, Tao, Degeling, Zong, Wang, Tian, {et~al.}}]{yao2019waves}
Yao, S., Shi, Q., Yao, Z., {et~al.} 2019, Geophysical Research Letters, 46, 523

\bibitem[{Yu {et~al.}(2021)Yu, Huang, Yuan, Jiang, Xiong, Xu, Wei, Zhang, \& Zhang}]{yu2021characteristics}
Yu, L., Huang, S., Yuan, Z., {et~al.} 2021, The Astrophysical Journal, 908, 56

\bibitem[{Zhang {et~al.}(2008)Zhang, Russell, Baumjohann, Jian, Balikhin, Cao, Wang, Blanco-Cano, Glassmeier, Zambelli, {et~al.}}]{zhang2008characteristic}
Zhang, T., Russell, C., Baumjohann, W., {et~al.} 2008, Geophysical Research Letters, 35

\bibitem[{Zhang {et~al.}(2017)Zhang, Artemyev, Angelopoulos, \& Horne}]{zhang2017kinetics}
Zhang, X.-J., Artemyev, A., Angelopoulos, V., \& Horne, R. 2017, Journal of Geophysical Research: Space Physics, 122, 10

\end{thebibliography}
\bibliographystyle{aasjournal}



\end{CJK*}
\end{document}